\documentclass[fleqn,10pt]{wlscirep}
\usepackage[utf8]{inputenc}
\usepackage[T1]{fontenc}
\usepackage{xcolor}
\title{Scientific Reports Title to see here}

\begin{document}

\title{Microwave-Assisted Tunneling in Hard-Wall InAs/InP Nanowire Quantum Dots}

\author[1,2]{Samuele Cornia}
\author[3]{Francesco Rossella}
\author[3]{Valeria Demontis}
\author[3]{Valentina Zannier}
\author[3]{Fabio Beltram}
\author[3]{Lucia Sorba}
\author[1,2]{Marco Affronte}
\author[2,*]{Alberto Ghirri}

\affil[1]{Dipartimento di Scienze Fisiche Informatiche e Matematiche, Universit\`a di Modena e Reggio Emilia, via G. Campi 213/A, 41125 Modena, Italy}
\affil[2]{Istituto Nanoscienze - CNR, via G. Campi 213/A, 41125 Modena, Italy}
\affil[3]{NEST, Scuola Normale Superiore and Istituto Nanoscienze - CNR, Piazza San Silvestro 12, 56127 Pisa, Italy}

\affil[*]{alberto.ghirri@nano.cnr.it}

\begin{abstract}
With downscaling of electronic circuits, components based on semiconductor quantum dots are assuming increasing relevance for future technologies. Their response under external stimuli intrinsically depend on their quantum properties. Here we investigate single-electron tunneling in hard-wall InAs/InP nanowires in the presence of an off-resonant microwave drive. Our heterostructured nanowires include InAs quantum dots (QDs) and exhibit different tunnel-current regimes. In particular, for source-drain bias up to few mV Coulomb diamonds spread with increasing contrast as a function of microwave power and present multiple current polarity reversals. This behavior can be modelled in terms of voltage fluctuations induced by the microwave field and presents features that depend on the interplay of the discrete energy levels that contribute to the tunneling process.
\end{abstract}

\flushbottom
\maketitle

\thispagestyle{empty}

\section*{Introduction}

Single (few) electron transistors based on semiconductor quantum dots (QDs) are flexible solid-state components characterised by extensive control of charge, orbital and spin degrees of freedom. Electrons fill the dot in a shell structure in analogy with artificial three dimensional atoms and their wavefunctions depend on the shape and size of the system
\cite{TaruchaPRL96, KouwenhovenScience97, BjorkNanoLett04}. Orbital properties, in turn, determine the tunneling current and more in general the QD response to external stimuli.
Bottom-up grown InAs nanowires (NWs) recently emerged as a reliable platform to produce single \cite{BjorkNanoLett04, FasthPRL08, RoddaroNanoLett, RomeoNanoLett} and double \cite{FuherNanoLett07, RossellaNatNano} QDs with strong electron confinement. In the case of high aspect ratio NW QDs, lowest lying states have dominant radial character while an axial component can  characterise their excited states \cite{BjorkNanoLett04}. Typically, the separation of their energy levels can be controlled between few tenths to tens of meV during growth \cite{SalfiReview}, while further tuning of NW QD energy levels can be obtained by (multi) gating \cite{RoddaroNanoLett}. Heterostructured InAs/InP NW QDs are characterised by hard-wall confinement potential and large single-particle energy spacing with Coulomb and Pauli blockade detectable up 50 and 10 K, respectively \cite{RoddaroNanoLett, RossellaNatNano}. Thanks to their large spin-orbit coupling, InAs NW QDs have been proposed as spin qubits with electric control of the spin degree of freedom \cite{TrifPRB08, Nadj-PergeNature10}.

QDs coupled to microwaves (MW) for spintronics and quantum technologies have been also proposed and developed \cite{AwschalomScience, MiScience17, StockklauserPRX17, SamkharadzeScience18, MiNature18, Loss2019}. When the photon energy $\hbar \omega$ matches the level spacing, coherent and/or resonant phenomena may occur and NW QDs may function either as single-atom maser sources \cite{Lambert, LiuPRL17, Mantovani} or as photon detectors in the MW range \cite{WongPRA}. 
Non-resonant MW excitation may assist tunneling process thus affecting the charge transport characteristics, as observed on other semiconductor quantum well\cite{BarbieriAPL} and QD systems: effects of electromagnetic radiation on Coulomb blockade peaks were reported for electrostatically defined GaAs QDs \cite{McEuenNano, Kouwenhoven, FreyPRB12} and single-walled carbon nanotubes \cite{IshibashiPhysB02, DelbecqNatComm}. The lifting of Coulomb blockade in the presence of MW radiation can be described in terms of photon-assisted tunneling (PAT) \cite{tien} that, in the quantum limit ($\hbar \omega >k_B T$), gives rise to inelastic single-electron tunneling \cite{KouwenhovenPRL94}. In the high MW power regime, sidebands of the main Coulomb peaks appear and excited states within the QD can contribute to assist the tunneling process \cite{Kouwenhoven}. This situation was not explored so far in heterostructured InAs/InP NW QDs despite their significant technological potential.   

Here, we first characterise the stability diagram of InAs/InP NW QDs; we then study Coulomb-blockade lifting in the presence of an off-resonant MW drive. We find that MW effects exhibit a power threshold and, under suitable conditions, current polarity reversal can be observed. Our results show that MW assisted tunneling can take place in non-resonant conditions as an effect of the voltage fluctuations of the microwave field and may depend on the presence of excited states of the QD involved in the tunneling process.

\section*{Results}

\subsection*{Device Description and Characterisation}

\begin{figure}[ht]
\centering
\includegraphics[width=\linewidth]{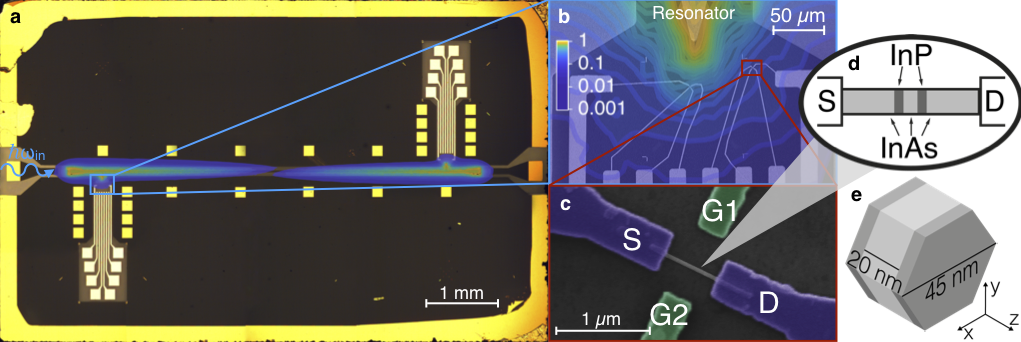}
\caption{(a) Optical microscope image of the whole electrical device with central YBCO/sapphire coplanar resonator. Gold contacts and bonding pads are used for connecting dc lines and the coplanar resonator to the external cables. (b) Scanning Electron Microscope (SEM) close-up showing the antenna tip and the leads of two NW QD devices. In panels (a) and (b) the finite-element simulation of the distribution of the electric component of the fundamental mode of the resonator is superimposed with false colours. The colour scale is normalized to the maximum  value. (c) False colour SEM image of the NW QD device investigated in this work, where source (S), drain (D) and gate (G1, G2) leads are indicated. (d) Schematic diagram of the InAs/InP NW QD. (e) Sketch of the NW QD showing InAs (light grey) and InP (dark grey) sections.}
\label{fig1}
\end{figure}

In order to optimise the coupling with microwaves, NW QD devices were fabricated in proximity of a half-wavelength superconducting coplanar resonator. The resonator was realised out of a $8 \times 5 $ mm YBCO/sapphire film and presents a 6 mm-long central conductor, having width of 30~$\mu$m and distance of 55~$\mu$m from the lateral ground planes. The fundamental mode has resonance frequency $\omega_0/2 \pi=9.815$~GHz (Supplementary Material). QD devices are fabricated in correspondence of the electric antinodes of the fundamental mode of the resonator, where two tips connected to the central strip are designed to work as "antenna", in order thus to extend and enhance the microwave field in proximity of the NW QD device (Fig.~\ref{fig1}(a,b)). Up to four QD devices can be individually tested for each chip. In this work we report results obtained on one of our NW QD devices, positioned at 65 $\mu$m from the antenna tip (Fig.~\ref{fig1}). 

\begin{figure}
\centering
\includegraphics[width=\linewidth]{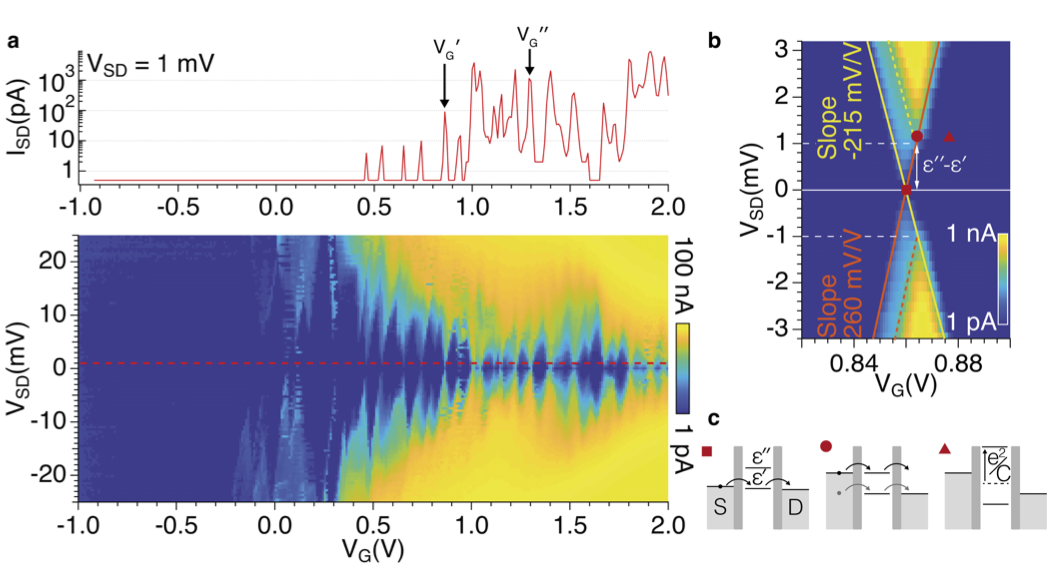}
\caption{(a) Stability diagram (bottom panel) and crosscut at $V_{SD}=1$~mV (top panel) of the $\log \left(|I_{SD}|\right)$ current measured at $T=2$~K. (b) Blow-up of the current map measured in the proximity of the peak at $V_G^{\prime} = 0.86$~V. The colour scale is logarithmic. (c) Schematic diagrams showing the configuration of the electrochemical potentials of ground ($\epsilon^{\prime}$) and first excited ($\epsilon^{\prime\prime}$) states in correspondence of the symbols in panel (b).}
\label{fig2}
\end{figure}

In order to characterise the QD we first measured the dc current ($I_{SD}$) as a function of bias voltage ($V_{SD}$) and gate voltage ($V_{G}$). The same voltage was applied simultaneously to both side gates ($V_{G1}=V_{G2}=V_G$). The stability diagram of the NW QD measured at the temperature $T=2$~K presents a typical Coulomb-diamond structure (Fig.~\ref{fig2}(a)). The addition energy of the $N$-th electron in the QD can be estimated in the framework of the constant interaction model \cite{HansonRMP}:
\begin{equation}
\mu_N = E_k +\frac{Ne^2}{C} -\vert e \vert \alpha V_{G} +c,
\end{equation}
where $E_k$ is the energy of the first available QD level, $C$ is the capacitance of the QD and $c$ is a constant. The lever arm $\alpha$ is a geometry-dependent parameter that accounts for the effect of gate voltage on QD levels, which can be obtained from the diamond boundary slopes $m_1$ and $m_2$ as $1/\alpha = 1/m_1 - 1/m_2$. Data in Fig.~\ref{fig2}(a) yield average values $\alpha = 120\pm 20$ mV/V and a charging energy $e^2/C= 9 \pm 2$ meV.  The split of the energy levels in our QD well matches values obtained for hard-wall InAs/InP NW QDs ($ca.$ 10 meV) having similar geometry and stronger axial confinement along the growth direction of the NW \cite{RoddaroNanoLett, SalfiReview}. 

The plot of $I_{SD}$ at constant voltage bias $V_{SD}=1$~mV is  shown in the top panel of Fig.~\ref{fig2}(a). As a function of the gate voltage, full depletion and progressive occupation of the QD states can be observed. Although collected data do not allow for a direct correspondence of current peaks with electron levels with spectroscopic precision, we expect that electrons are added to fill in the shell structure of the QD, which is determined by the specific characteristics of the electron orbitals \cite{TaruchaPRL96, KouwenhovenScience97, BjorkNanoLett04}.

Data plotted in log-scale in Fig.~\ref{fig2}(a) show a marked increase of the peak current from $\sim 10$ pA to $\sim 1$ nA beyond a threshold voltage of 1~V. These abrupt change reflects the different tunneling rates ($\Gamma$) that can be estimated by considering the Fermi-Dirac occupation of the source and drain leads at $T=2$~K (Supplementary Material). By fitting the $I_{SD}(V_{SD})$ peaks, we obtain $\Gamma^{\prime}=1$~GHz  for $V_G^{\prime}=0.86$~V, whilst for $V_G>1$~V we estimate $\Gamma^{\prime\prime} \approx 10$~GHz. These numbers are consistent with what previously observed in similar QD devices \cite{RoddaroNanoLett}. 

The stability diagram taken as function of the gate voltage for $V_G \sim V_G^{\prime}$ is shown in Fig.~\ref{fig2}(b). The diamond exhibits two different slopes ($m_1=-215$~mV/V and $m_2=260$~mV/V) that correspond to slightly asymmetric lever arms for source and drain leads. 
Fig.~\ref{fig2}(b) shows that $I_{SD} \propto \Gamma^{\prime} V_{SD}$ for voltage gate $V_G=V_G^{\prime}$ and bias $\left |V_{SD} \right |<1$~mV, whilst a steeper dependence $I_{SD} \propto \Gamma^{\prime \prime} V_{SD}$ is observed for $\left |V_{SD} \right |>1$~mV. This behavior suggests that the Coulomb peak is related to tunneling through two charge states with the presence of an excited level at energy $\Delta E_k =\epsilon^{\prime \prime}-\epsilon^{\prime} \approx 1$~meV. 

The electron tunneling through the QD can be depicted as in Fig.~\ref{fig2}(c), which shows the configuration of the electrochemical potentials $\mu(\epsilon^{\prime})$ and $\mu(\epsilon^{\prime\prime})$ as a function of the gate voltage around $V_G^{\prime}$. For $V_G=V_G^{\prime}$ and $V_{SD}=1$~mV (square symbol in Fig.~\ref{fig2}(b,c)), the tunneling involves only the energy level $\epsilon^{\prime}$, thus the low tunneling rate $\Gamma^{\prime}$ gives rise to a low tunneling current. By expanding the bias window ($\left |V_{SD} \right |>1$~mV) (circle symbol), the transmission channel through $\epsilon^{\prime\prime}$ becomes accessible and contributes to the electron transport. A sudden increase of the tunneling current is observed as a consequence of the larger tunneling rate $\Gamma^{\prime\prime} $. When $\left | V_G - V_G^{\prime} \right |$ is sufficiently large (triangle symbol), the level $\epsilon^{\prime}$ is permanently filled and the addition of an extra electron costs the charging energy $e^2/C$. In this case, $\epsilon^{\prime\prime}$ is no longer accessible and the current flow is blocked.

\begin{figure}
\centering
\includegraphics[width=\linewidth]{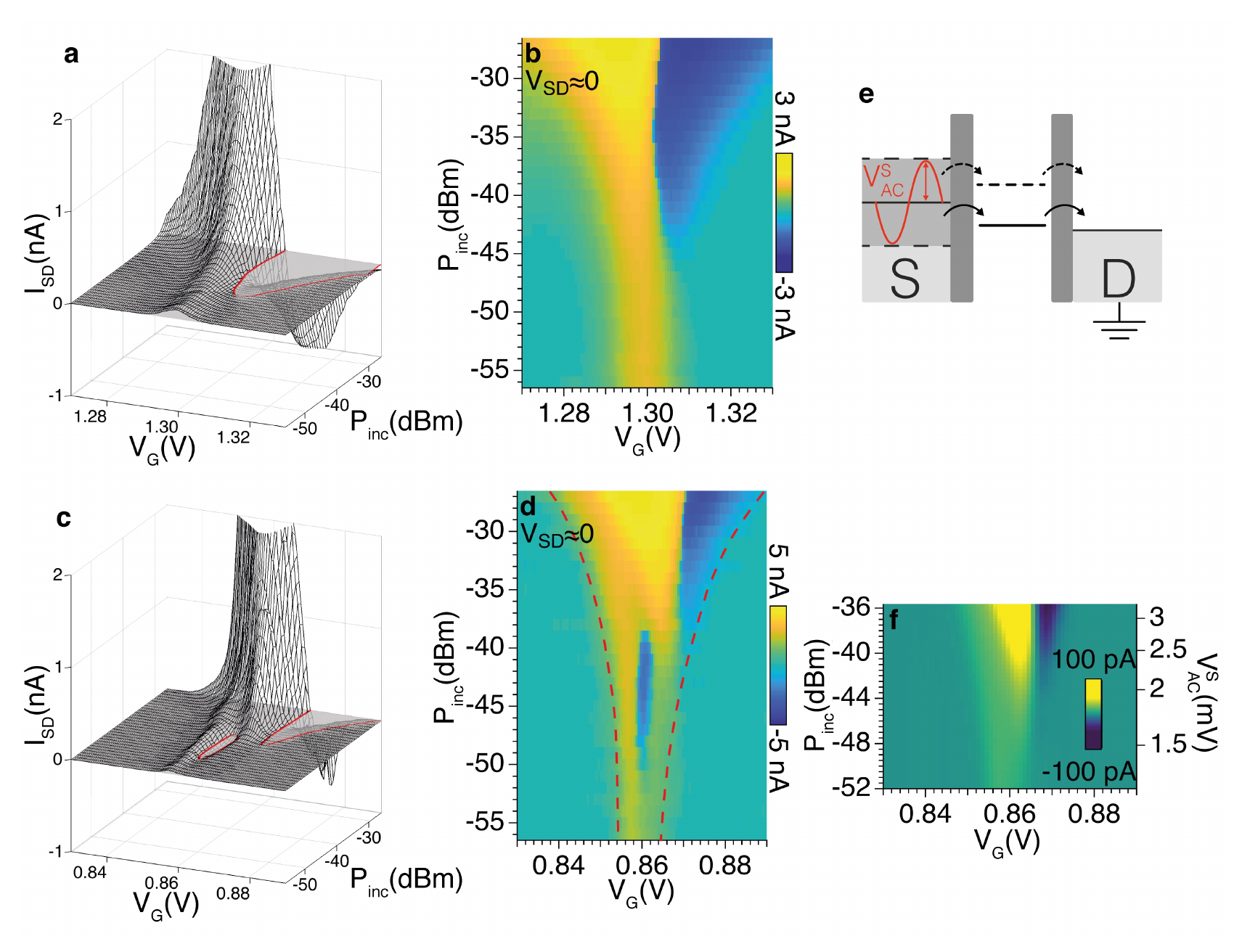}
\caption{Evolution of the $I_{SD}(V_G)$ characteristics in the presence of a microwave drive of frequency $\omega_0$ and increasing power $P_{inc}$. Three dimensional plots and maps are  measured for gate voltage around $V_G^{\prime \prime}=1.30$~V (a,b) and $V_G^{\prime}=0.86$~V (c,d) at the temperature $T=2$ K. Solid lines indicate the contour of regions with negative $I_{SD}$. The dashed line in panel (d) display the zero current points used to extract the peak width $\Delta V_G$. (e) Schematic energy diagrams showing the MW-assisted tunneling through the dot levels $\epsilon^{\prime}$ and $\epsilon^{\prime \prime}$. (f) $I_{SD}(V_G)$ characteristics calculated for $V_{SD}\simeq 0$ by averaging the measured current over increasing voltages $V_{AC}^S$.}
\label{fig3}
\end{figure}

\subsection*{Microwave-Assisted Transport}

In order to study the effect of the microwaves on the transport properties of InAs/InP NW QDs, current measurements were performed in the presence of a monochromatic wave ($\omega_0/2 \pi=9.815$~GHz). In the following we shall focus on two Coulomb peaks at $V_G^\prime=0.86$~V and $V_G^{\prime \prime}=1.30$~V that are representative of the two electron transport regimes as previously described. We initially focused on the Coulomb peak at $V_G^{\prime \prime} = 1.30$~V to investigate the $I_{SD}(V_G)$ characteristics by varying the microwave power (Fig.~\ref{fig3}(a,b)).  As a general observation, the presence of an intense MW drive leads to amplification of the tunnel current and, as expected, this is particularly evident close to $V_{SD}=0$. No effect on electron transport is detected for $P_{inc} <-55$~dBm, while MW impact increased more than linearly with increasing MW power. For $P_{inc}>-45$~dBm a polarity reversal of $I_{SD}$ and the broadening of the Coulomb peak are visible. In particular, a negative current peak appears for $V_G>1.30$~V even with $V_{SD}>0$. Such trends are found for all Coulomb-blockade peaks that were investigated. 

In the case of the Coulomb peak at $V_G^{\prime} = 0.86$~V (Fig.~\ref{fig3}(c,d)), an additional dip with $I_{SD}<0$ ($i.e.$ current polarity reversal) is present in the intermediate power range ($-50<P_{inc}<-38$~dBm) and steps appear in the $I_{SD}$ characteristics  for $P_{inc}>-38$ dBm. 
In order to further investigate this point, we mapped $I_{SD}$ as a function of $V_{SD}$ and $V_G$ under constant MW power. Figure~\ref{fig4}(a-c) shows the evaluation of the Coulomb diamond with increasing $P_{inc}$. For $P_{inc}=-36.5$ dBm two current peaks are visible in panel (c). This effect is enhanced for high power (panel (d)), where the two peaks broaden and merge. The additional dip in Fig.~\ref{fig3}(c,d) for -50~dBm~$<P_{inc}<$~-38~dBm can thus be related to the evolution of the diamond shape as a function of the MW power, which gives rise to multiple reversal of the current in the $I_{SD}(V_G)$ characteristics for $V_{SD}\approx 0$.

In hard-wall InAs/InP NW QDs Coulomb blockade persists up to $T \approx 50$ K \cite{RoddaroNanoLett}, thus the effect of the MW field on the transport characteristics can be tested at intermediate cryogenic temperature. At 8 K the Coulomb peaks are much broader, yet the reversal of $I_{SD}$ is still visible for large power levels (See Supplementary Material). The comparison between 2 and 8 K data shows that the effect of the temperature on the $I_{SD} (V_G)$ characteristics is qualitatively different from that of the microwave drive, ruling out the possibility that the observed behavior is simply due to heating. 

\section*{Discussion}

\begin{figure}
\centering
\includegraphics[width=\linewidth]{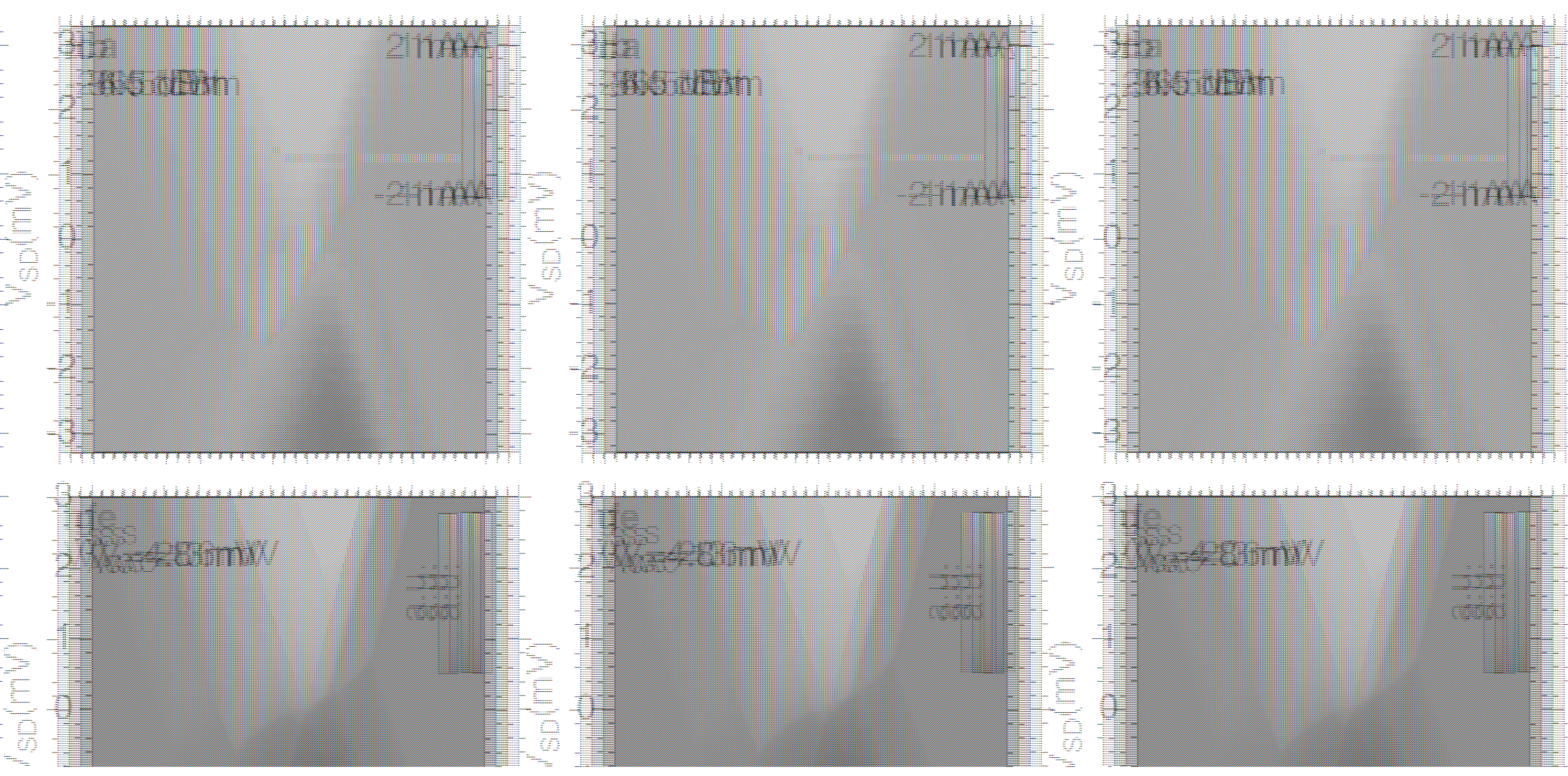}
\caption{(a,b,c,d) Charge stability diagrams measured around the Coulomb peak at gate voltage $V_G^{\prime}$=0.86~V.  (e,f,g,h) Calculated evolution of the stability diagrams in presence of different voltages $V_{AC}^S$ as reported in the text.}
\label{fig4}
\end{figure}

In NW QDs used in our experiments, the separation between the energy levels of the QD is orders of magnitude larger than the energy of the fundamental mode of the resonator ($\hbar \omega_0 \approx 40$ $\mu$eV). This implies that the observed effects (broadening of the Coulomb peak and reversal of the tunneling current) are not related to resonant transitions between single-particle energy levels. On the other hand, the effects of MW are evident for finite intensity of applied electromagnetic radiation:
in the many-photon regime and for $k_B T > \hbar \omega_0$, the microwave mode can be treated as a classical electromagnetic field that induces a broadening of the electrodes energy levels \cite{McEuenNano, Kouwenhoven, IshibashiPhysB02, DelbecqNatComm}. More specifically, the coupling with the microwave field induces an oscillating voltage of amplitude $V_{AC}^S$ on the source lead. At $V_{SD}=0$, and for a given $V_G$, the current flows when $V_{AC}^S$ matches the equivalent value of $\left|V_{SD} \right|$ that would lift the Coulomb blockade. Therefore the width of the peak results $\Delta V_G = 2  V_{AC}^S /\min \left( \left| m_1 \right|  , \left| m_2 \right| \right)$, where $m_1$ and $m_2$ are the slopes of the diamond edges. For a given power $P_{inc}$, $V_{AC}^S$ can be estimated from the broadening of the Coulomb peak in Fig.~\ref{fig3}(d). We calculated the $I_{SD}(V_G)$ characteristics by following the method suggested in Refs.~\cite{IshibashiPhysB02, DelbecqNatComm} which consists in calculating, for each $V_G$ value, the average of the current over the $\pm V_{AC}^S$ interval. Curves obtained from the experimental data in Fig.~\ref{fig2}(b) are displayed in Fig.~\ref{fig3}(f). It is worth noting that the main trends of the experimental $I_{SD}(V_G)$ data shown in Fig. \ref{fig3}(a-d), in particular current amplification, peak broadening and reversal of current polarity for increasing voltage $V_{AC}^S$, are reproduced by this simple average calculation. Within the framework of this classical model, the reversal of current polarity emerges as a result of the asymmetry of the Coulomb diamonds ({\it i.e.} $m_1 \ne m_2$). 

More careful inspection of experimental spectra reveals however a more complex behaviour such as a multiple reversal of current polarity, as shown in Fig.~\ref{fig3}(c,d) and in the maps in Fig.~\ref{fig4}(a-d) and this demands more details. Here it is worth pointing out that multiple reversal of current polarity was observed in coincidence of the Coulomb peak at $V_G^{\prime}$ (Fig.~\ref{fig2}(b)), for which the presence of an additional excited states is evident in the measured stability diagram. Kouwenhoven et al.~\cite{Kouwenhoven} investigated the process of photon-assisted tunneling with spectroscopic resolution and suggested a model in which excited QD states also contribute to the tunneling. 
The phenomenological description of stability diagram we previously discussed allows us to extract the essential parameters to describe the effects of classical microwave field also in the presence of excited states in the QD. We firstly reproduce the main features of the DC stability diagram by means of the experimental parameters $m_1$, $m_2$, $\epsilon^{\prime\prime}-\epsilon^{\prime}$, $\Gamma^{\prime}$ and $\Gamma^{\prime\prime}$ and we assume $I_{SD} \propto\Gamma V_{SD}$ (Fig.~\ref{fig4}(e)). Then, to reproduce the evolution of the stability diagram for increasing power $P_{inc}$, we apply the average method reported above at finite bias $V_{SD}$. In this way, we can calculate the average of the current for increasing $V_{AC}^S$ voltages. The outcome of the simulations (Fig.~\ref{fig4}(f-h)) well reproduces the main trends of the experimental data (Fig.~\ref{fig4}(b-d)).

To check the consistency of this phenomenological model, we compare the $V_{AC}^S$ voltages extracted from the measurements in Fig.~\ref{fig3} with the voltage fluctuation generated by the microwave field ($V_{AC}$). At finite $P_{inc}$, we calculate the root mean square voltage fluctuation as $V_{AC}=\sqrt{2} V_{AC}^{zpf} \sqrt{n+1/2}$, where  $V_{AC}^{zpf}=2.3$~$\mu$V is the zero point fluctuation in the center the coplanar resonator and $n$ is the number of photons (Supplementary Materials). We introduce the effective coupling factor $\lambda = 5 \times 10^{-2}$ such as $V_{AC}^S =\sqrt{2} \lambda V_{AC}$. Finite element simulations indicate that the value of $\lambda$ is consistent with the estimated decrease of the root mean square electric field between the electric antinode and the position of the QD (Fig.~\ref{fig1}(b) and Supplementary Material). We also expect that the misalignment between the direction of the electric field and the NW axis contributes in determining its value. 

In conclusion, we have investigated the influence of MW radiation on the transport characteristics of InAs/InP nanowire QDs and found that, above a threshold power, the MW field induces a broadening of the current peak and a current polarity reversal. These effects are relevant for microwave-assisted tunneling processes in the many-photon regime and reflect the discrete nature of QDs excited states. From these experiments we learn that a suitable choice of the working point, as well as of the electron wavefunction, enables one to control the tunneling current in strongly confined NW QD systems under external MW excitation.

\section*{Methods}

The superconducting coplanar resonator has been fabricated by optical lithography starting from commercial (Ceraco GmbH) Au(200~nm)/YBCO(330~nm)/sapphire(430~$\mu$m) multilayer films. Excess Au and YBCO regions were etched by argon plasma in a reactive ion etching (RIE) chamber. In this process, gold pads were defined on top of the YBCO film in order to serve as wire bonding spots and used either to link the coplanar launchers to the external feedlines or for grounding connections. Additional leads and pads (eight per each slot) for dc measurements were fabricated by lift-off of thermally evaporated Au(100~nm)/Ti(10~nm)/sapphire films.

InAs nanowires with hexagonal section and $45 \pm 5$ nm diameter were grown by metal-assisted chemical beam epitaxy \cite{ZannierNanotech19}. By changing the precursor from  tert-butylarsine to tert-butylphosphine during NW growth, two $5 \pm 1$ nm layers of InP separated by $20 \pm 1$ nm of InAs were obtained. These InP layers act as tunneling barriers and define the InAs QD in the NW (Fig.~\ref{fig1}(d)). After growth, NWs were detached from the InAs substrate by sonication in isopropyl alcohol and drop cast onto a sapphire substrate where the superconducting resonator had been previously realised. Ohmic contacts between either source (S) or drain (D) electrodes and NW were obtained by e-beam lithography followed by chemical passivation in NH$_4$S$_x$ solution, thermal evaporation of Au(100 nm)/Ti(10 nm) films and lift-off. Additionally, two local gate electrodes (G1 and G2) were fabricated in the proximity of the NW (Fig.~\ref{fig1}(c)). The two nanogates are placed in the middle of the gap between S and D electrodes at a distance of approximately 400 nm from the NW QD. Electron transport measurements were carried out with the S electrode connected to voltage $V_{SD}$ and the D electrode at ground potential. The latter was connected to a transconductance amplifier with gain in the $10^7-10^9$ range for dc current measurements. No backgate bias was applied. Continuous wave microwave tones were generated by a fixed-frequency microwave source. The incident power is estimated at the input of the resonator, feedline attenuation was taken into account.

\section*{Acknowledgments}
We thank S. Roddaro for useful discussions and D. Prete for the assistance during the experiments. This work was partially supported by CNR-NANO Seed Research Award 2016, by the Air Force Office of Scientific Research grant (contract no FA2386-17-1-4040) and by the SUPERTOP project, QUANTERA ERA-NET Cofound in Quantum Technologies. This is a post-peer-review, pre-copyedit version of an article published in Scientific Reports. The final authenticated version is available online at: https://doi.org/10.1038/s41598-019-56053-2 .

\section*{Author Contributions Statement}
A.G. and F.R. conceived the experiment. S.C. and A.G.  carried out the measurements and analyzed the data with inputs from M.A. and F.R. V. Z. and L.S. carried out the nanowire growth process. V.D. and S.C. carried out the device fabrication. S.C., A.G., M.A., F.R. and F.B. cowrote the manuscript with contributions from all authors.

\section*{Additional Information}
The authors declare no competing interests.

\section*{Supplementary Material}

\subsection*{Electromagnetic Characterisation}

\begin{figure}[b]
\centering
\includegraphics[width=15 cm]{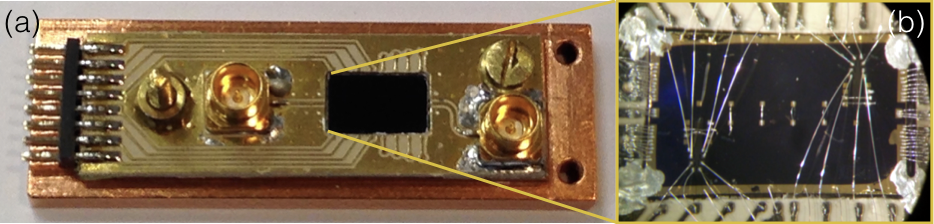}
\caption{(a) Gold-plated PCB realised on a high dielectric constant AD1000 laminate. Microwave lines are connected through SMP connectors to the coaxial cables. dc lines are linked to the 16 pin connector positioned on the left side of the PCB. The board is installed into an oxygen free high conductivity copper box. (b) Device bonded on the PCB board.}
\label{photo_device}
\end{figure}

\begin{figure}
\centering
\includegraphics[width=15 cm]{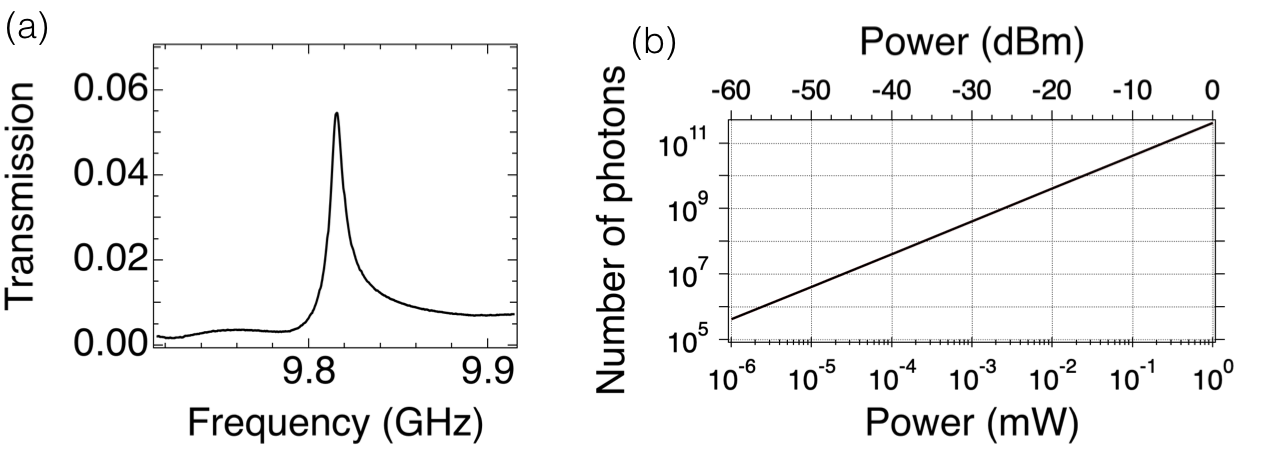}
\caption{(a) Transmission spectrum of the coplanar resonator showing the fundamental mode ($T=4$ K). The incident microwave power is $P_{inc}=-13$ dBm. (b) Average photon population as a function of $P_{inc}$.}
\label{figure_resonance}
\end{figure}

Electrical characterisation and measurements under microwave radiation were performed down to 2~K in a Quantum Design PPMS by means of a low temperature probe wired with 16 dc filtered lines and 2 coaxial cables. The hybrid device was enclosed in a copper box, in which microwave and dc lines were wire bonded to a gold-plated printed-circuit board (PCB) (Fig.~\ref{photo_device}). A stage of attenuators was inserted at low temperature to suppress the heat radiated from room temperature components. The half-wavelength coplanar resonator was capacitively coupled through 140 $\mu$m wide gaps to the coplanar launchers located on the short sides of the chip that, in turns, connect the resonator to the external feedlines. 

The transmission spectrum in Fig.~\ref{figure_resonance}(a) shows the fundamental mode of the YBCO/sapphire coplanar resonator with resonance frequency $\nu_0=\omega_0/2 \pi=9.815$~GHz. Typical quality factor of our bare YBCO resonators are of the order of $10^4$, while the loaded quality factor, as measured for the resonator in the configuration of the hybrid device including metal contacts, is $Q_L \approx 1500$. The insertion loss is $IL=25$~dB. To estimate the average photon population in the fundamental mode ($n$) we use the conventional formula\cite{SageJAP11}
\begin{equation}
n=\frac{1}{\pi h \nu_0^2 }P_{inc}Q_L10^{-IL/20}, 
\end{equation}
where $h$ is the Planck constant and $P_{inc}$ is the incident power. For the typical $P_{inc}$ values used in our experiment we estimate a photon number in the $10^{6}-10^{9}$ range (Fig.~\ref{figure_resonance}(b)).
\begin{figure}
\centering
\includegraphics[width=15 cm]{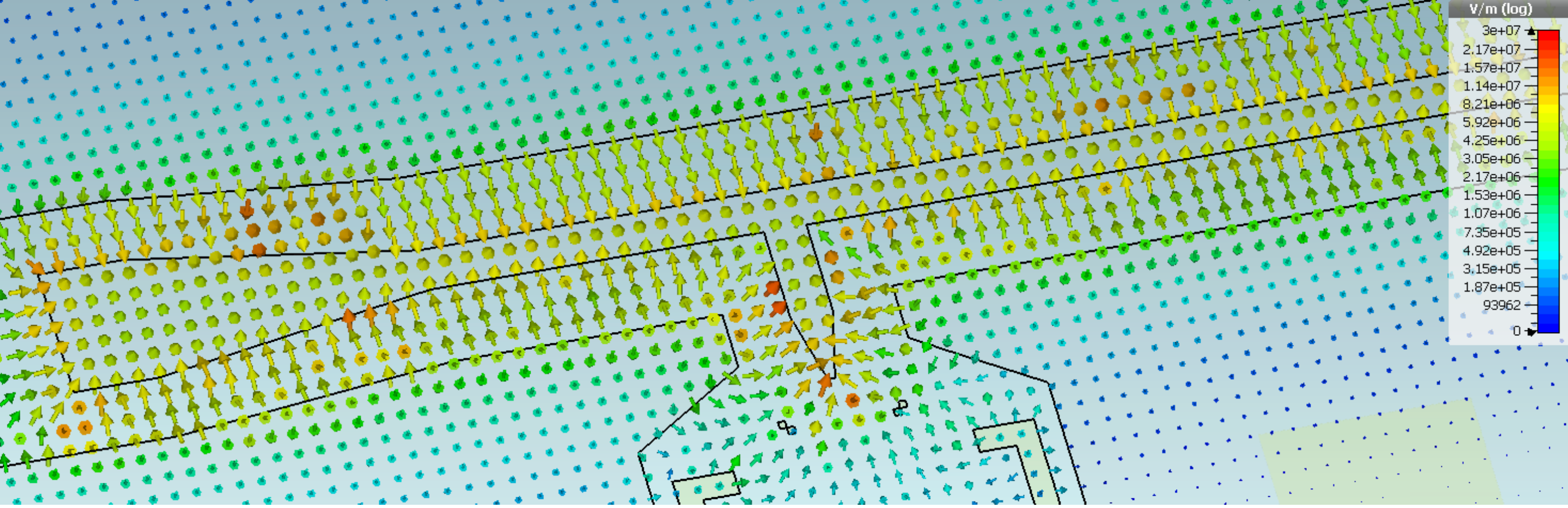}
\caption{Finite element simulation of the distribution of the electric field component of the fundamental mode of the resonator. The color scale shows the magnitude of the electric field for an incident microwave power $P_{inc}=1$~W.}
\label{simulation}
\end{figure}

Finite element simulations of the coplanar resonator were carried by means of a commercial software (CST Microwave Studio). Fig.~\ref{simulation} shows the simulated distribution of the electric field component of the fundamental mode of the resonator. From the simulation, the root mean square (rms) amplitude of the electric field at the antinode can be estimated as $E_{AC} \textrm{[V/m]}= 3.0 \times 10^7 \sqrt{P_{inc} \textrm{[W]}}$. To calculate the zero-point fluctuation, we consider an incident power corresponding to the vacuum state (-150~dBm). The rms amplitude of the vacuum fluctuation of the electric component results $E_{AC}^{zpf} \approx 3 \times 10^{-2}$~V/m. 

For comparison, the root mean square value of the vacuum fluctuation on the center conductor of the resonator can be estimated as $V_{AC}^{zpf}=\sqrt{\hbar \omega_0/2C_{res}}$ (Ref. \cite{WalraffNature04}). $C_{res}$ is the capacitance of the resonator, which can be calculated by analytical techniques and in our case results 0.6~pF.\cite{Lancaster} We thus obtain $V_{AC}^{zpf}=2.3$~$\mu$V. By considering that the distance between the center of the coplanar resonator and the ground planes is $w/2+s$, being $w=30$~$\mu$m the width of the central strip and $s=55$~$\mu$m the longitudinal separation between central strip and ground planes, we get $E_{AC}^{zpf}=3 \times 10^{-2}$~V/m. This value is in excellent agreement with the outcome of finite element simulations.

From the simulated distribution of $E_{AC}$ (Fig.~\ref{simulation}), we can estimate the rms amplitude at the position of the NW QD device as $E_{AC}^{\prime} \textrm{[V/m]} \approx 2 \times 10^5 \sqrt{P_{inc} \textrm{[W]}}$. This rms value is two orders of magnitude lower than that estimated at the electric antinode. For the range of incident power $-60 \textrm{~dBm}<P_{inc}<-20\textrm{~dBm}$ used in our experiments, the rms amplitude of the electric field results $10~\textrm{V/m}<E_{AC}^{\prime}<100~\textrm{V/m}$. We note that the effective length calculated from the ratio $V_{AC}^{lead}/E_{AC}^{\prime} \sim 30$ $\mu$m suggests that the electromagnetic wave effectively couples to the leads to modulate their voltage.

\subsection*{Tunneling Rate}

\begin{figure}[b]
\centering
\includegraphics[width=10 cm]{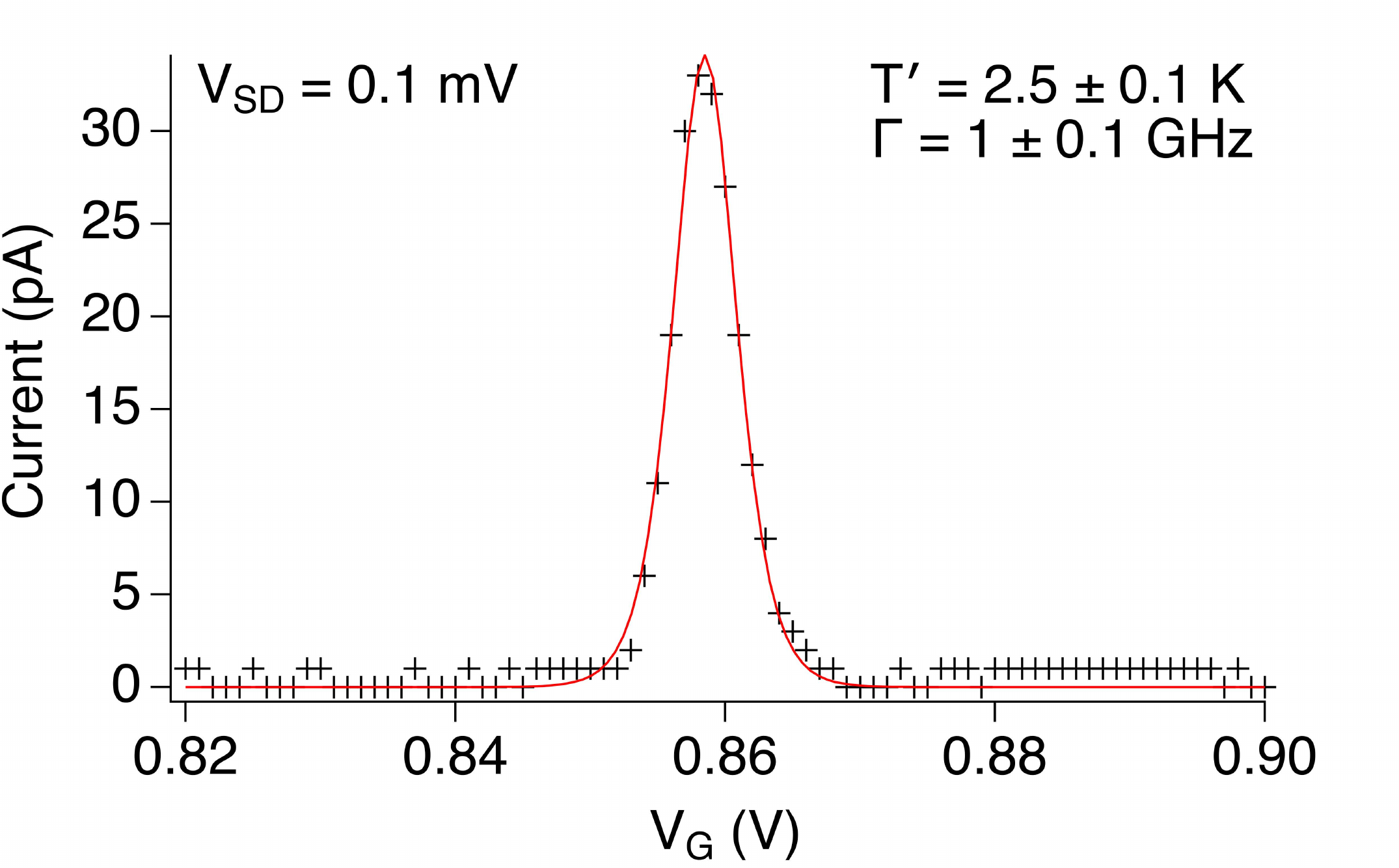}
\caption{Current $I_{SD}$ measured as a function of the gate voltage $V_G$ at $V_{SD}=0.1$~meV (symbols). The fit with Eq. \ref{eq:fit_coulomb} is displayed by the solid line.}
\label{tunneling_rate}
\end{figure}

The tunneling rate of the QD at $V_G^{\prime}=0.86$~V was determined by fitting the Coulomb peak measured in the low bias limit with the equation \cite{Houten}
\begin{equation}
G=\frac{I_{SD}}{V_{SD}} = \frac{e^2\Gamma^{\prime}}{k_BT^{\prime}}\frac{1}{4\cosh^2\left(\frac{e\alpha\left(V_G^{\prime}-V_G\right)}{2k_BT^{\prime}}\right)},
\label{eq:fit_coulomb}
\end{equation}
where $\alpha$ is the lever arm of the gates whilst $e$, $h$ and $k_B$ are respectively the elementary charge, the Planck and the Boltzmann constants. From the experimental data (Fig.~\ref{tunneling_rate}) we extracted $\Gamma^{\prime}=1.0 \pm 0.1$ GHz and $T^{\prime}=2.5 \pm 0.1$~K. The latter is compatible with the measured temperature of the experiment ($T=2$~K).

For Coulomb peaks at $V_G>1$~V the tunneling rate was simply estimated by the maximum value of the current at fixed bias.

\begin{figure}
\centering
\includegraphics[width=8 cm]{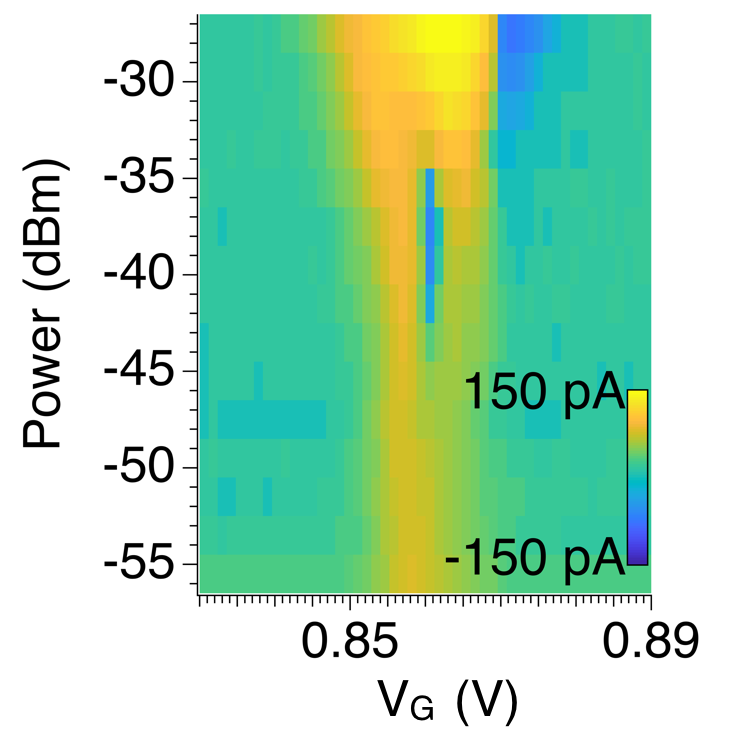}
\caption{Evolution of the $I_{SD}(V_G)$ characteristics in the presence of a microwave tone of  increasing power $P_{inc}$ and frequency $\omega_1/ 2 \pi = 9.810$~GHz. The temperature is $T=2.4$ K.}
\label{off_res}
\end{figure}

\subsection*{Additional Measurements}

Fig. \ref{off_res} shows the evolution of the Coulomb peak at $V_G^{\prime}$ for increasing values of $P_{inc}$ (Fig.~\ref{figure_resonance}). In this case the frequency of the microwave tone is $\omega_1 / 2 \pi = 9.810$~GHz. $I_{SD}(V_{SD})$ characteristics measured for increasing power $P_{inc}$ show features comparable to those reported in the main article for $\omega_0/2 \pi=9.815$~GHz. The vertical shift of the measured curves in terms of incident microwave power follows as a consequence of the reduced transmission at $\omega_1$. No significant frequency dependence was observed within the resonator bandwidth.

\begin{figure}
\centering
\includegraphics[width=10 cm]{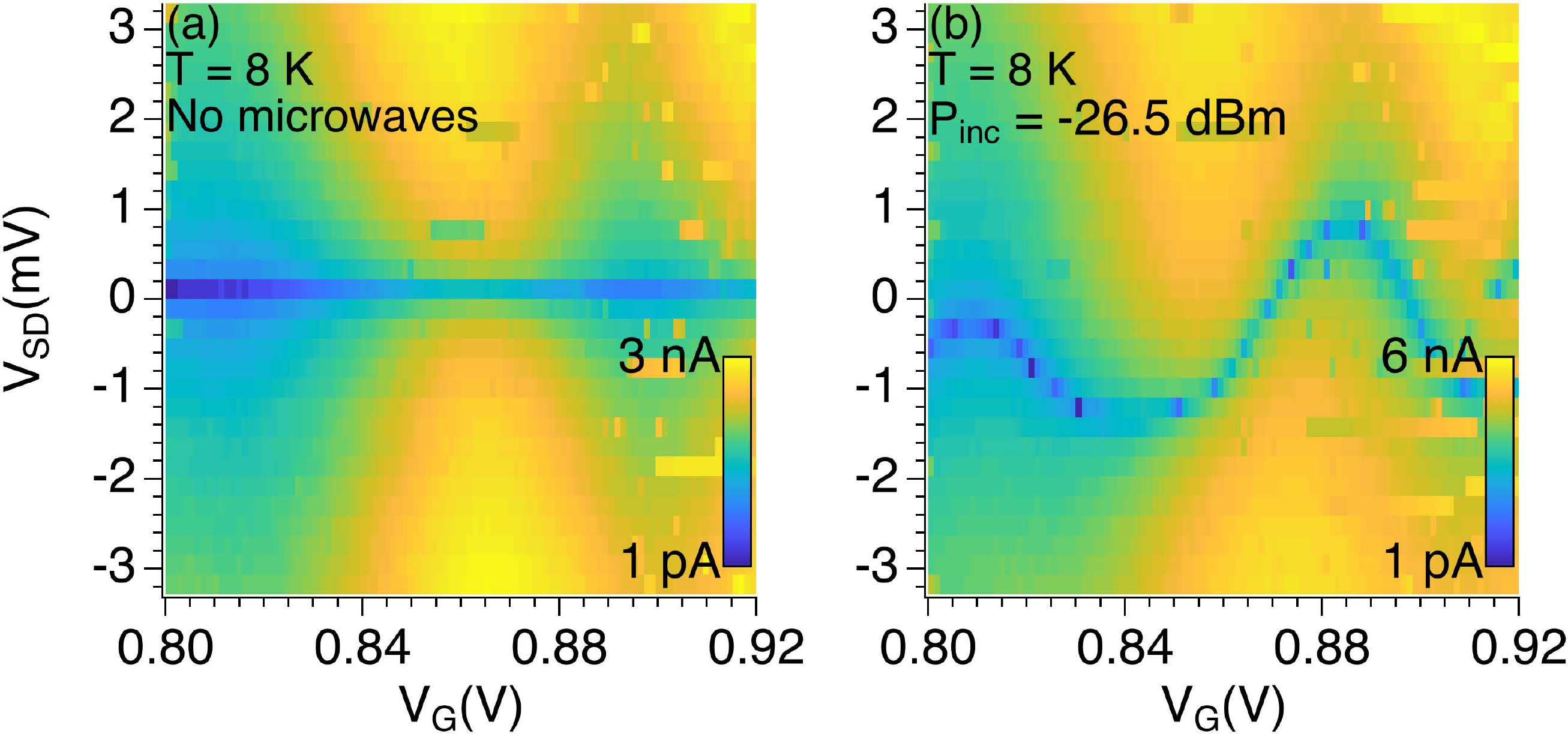}
\caption{Current ($I_{SD}$) maps taken at the temperature $T=8$~K. (a) $P_{inc}=0$, (b) $P_{inc}=-26.5$ dBm and $\omega=\omega_0$. }
\label{T8K}
\end{figure}

Fig. \ref{T8K}(a) shows the stability diagram measured near $V_G^{\prime}=0.86$ V at the temperature $T=8$ K. Respect to the experimental data taken at $T=2$ K (main article), the Coulomb peaks are broader due to the higher temperature. Current polarity reversal and peak broadening are visible in the presence of the microwave drive (Fig. \ref{T8K}(b)).

\bibliography{biblio}

\end{document}